\documentclass[twocolumn]{jpsj3}
\usepackage{graphicx}
\usepackage{bm}
\usepackage{txfonts}

\title{Face centered cubic SnSe as a $\mathbb{Z}_2$ trivial Dirac nodal line material }

\author{Ikuma Tateishi and Hiroyasu Matsuura}
\inst{Department of Physics, The University of Tokyo, Hongo, Bunkyo-ku, Tokyo 113-0033, Japan} 

\abst{The presence of a Dirac nodal line in a time-reversal and inversion symmetric system is dictated by the $\mathbb{Z}_2$ index when spin-orbit interaction is absent. 
In a first principles calculation, we show that a Dirac nodal line can emerge in $\mathbb{Z}_2$ trivial material by calculating the band structure of SnSe in a face centered cubic lattice as an example. We qualitatively show that it becomes a topological crystalline insulator when spin-orbit interaction is taken into account.
We clarify the origin of the Dirac nodal line by obtaining irreducible representations corresponding to bands and explain the triviality of the $\mathbb{Z}_2$ index.
We construct an effective model representing the Dirac nodal line using the {\bf k}$\cdot${\bf p} method, and discuss the Berry phase and a surface state expected from the Dirac nodal line.    
}

\begin{document}
\maketitle

Electronic states protected by a combination of crystal symmetries and topology have attracted much interest in the field of material science.
A Dirac electron system in a material is one of these protected electronic states, and its electronic state has been studied using first principles calculations and many experiments.
Recently, a new kind of semimetallic state called a Dirac nodal line has attracted attention, in which massless Dirac points (or nodes) form a continuous line in the energy-momentum space across the Fermi energy. 
For example, the existence of a Dirac nodal line has been suggested in Cu$_3$NPd\cite{1}, CaTe\cite{2}, Ca$_3$P$_2$\cite{3}, and Ca\cite{3.05} based on first principles calculations.
Experimentally, it has been confirmed in some materials using photoemission spectroscopy \cite{3.1,3.2}. 
The properties of a Dirac nodal line without spin-orbit interaction are related to those with spin-orbit interaction\cite{4}. Therefore, understanding Dirac nodal lines without spin-orbit interaction can contribute to an understanding of the topological properties of systems with spin-orbit interaction.

Generally, a Dirac nodal line lies on general points in the Brillouin zone (BZ).
Thus, in order to find the Dirac nodal line, it is necessary to check the entire BZ using a first principles calculation.
However, this is costly in terms of time. 
Therefore, an alternative method that uses only the information of high symmetry points is desirable.
For systems with time-reversal (TR) symmetry and inversion symmetry, a method using the $\mathbb{Z}_2$ index was proposed\cite{1}. [This $\mathbb{Z}_2$ index is the same as that of topological insulators with TR and inversion symmetry\cite{5}.]
In general, when a system has the TR and inversion symmetry and does not have spin-orbit interaction, a point node (or Dirac point) cannot exist\cite{15}. However, when such a system has a non-trivial $\mathbb{Z}_2$ index, it has been shown that the Dirac nodal line exists, and the system is semimetallic.

The detailed relationship between the $\mathbb{Z}_2$ index and the Dirac nodal line is as follows.
When spin-orbit interaction is absent, the $\mathbb{Z}_2$ index $\nu_i$ is obtained by $(-1)^{\nu_i} = \exp ( i \gamma_i )$, where $\gamma_i$ is the Berry phase defined on a TR invariant loop that goes through four specific time-reversal invariant momenta (TRIM).
When this TR invariant loop is penetrated by a single nodal line, the Berry phase is $\pi$.
In the same way, when the TR loop is penetrated by two nodal lines, the Berry phase becomes $2\pi$, which results in a trivial $\mathbb{Z}_2$ index. There are many materials with Dirac nodal lines protected by non-trivial $\mathbb{Z}_2$ indices \cite{4}. The topological properties of these systems have been discussed for the case where spin-orbit interaction is taken into account \cite{4}. On the other hand, there are few materials with Dirac nodal lines and trivial $\mathbb{Z}_2$ indices. Ca \cite{3.05} is one of these materials. However, the topological properties of the system have not yet been discussed for the case where spin-orbit interaction is taken into account.

In this letter, we show new explicit examples of systems with trivial $\mathbb{Z}_2$ indices and Dirac nodal lines. 

It is worth noting the effect of spin-orbit interaction. When the spin-orbit interaction is taken into account in the above systems with Dirac nodal lines, it has not been clarified what kinds of states are realized. However, it will be natural to expect that a $\mathbb{Z}_2$ trivial semimetal or insulator would be realized. In the following, we will show that there are some examples where a Dirac nodal line system becomes a $\mathbb{Z}_2$ trivial insulator, and particularly a topological crystalline insulator in the presence of spin-orbit interaction. It has been shown that a topological crystalline insulator\cite{6} has a trivial $\mathbb{Z}_2$ index but a topological non-trivial gapless surface state.
Their surface states are protected by their crystal symmetries such as mirror symmetry.
One well known example of a topological crystalline insulator is SnTe\cite{7,8}, which has an inverted band structure and strong spin-orbit interaction.
Therefore, when the spin-orbit interaction is absent, SnTe is a candidate material with a Dirac nodal line. The related material SnSe is also a candidate. SnSe has been theoretically suggested to be a topological crystalline insulator\cite{8.5}. However, for a case without spin-orbit interaction, the topological properties of the system have not been well studied, especially from the perspective of a Dirac nodal line.
In the following, we focus on SnTe and its related material SnSe.
We use first principles calculations to determine the band dispersions of SnTe and SnSe. As a result, we show that Dirac nodal lines exist in $\mathbb{Z}_2$ trivial face centered cubic (f.c.c.) SnSe.
We construct an effective model representing the Dirac nodal line based on the {\bf k}$\cdot${\bf p} method, and discuss the Berry phase and a surface state expected from the Dirac nodal line. 

 All of the following calculations are performed using Quantum ESPRESSO\cite{qe}, which uses the plane wave density functional theory.
For the exchange correlation term, generalized gradient approximation (GGA) with non-relativistic Perdew-Burke-Ernzerhof\cite{ 52} parametrization is used.

First, we calculate the band structure of SnTe. For the lattice constant, we use $a_0 = 6.313 ~\mathbb{\AA}$, which was used in the previous study.\cite{8.1}. We find that band inversion does not occur in SnTe without spin-orbit interaction. When the spin-orbit interaction is taken into account, the band energies are shifted, and the band inversion takes place. However, we do not study SnTe further because we wish to focus on a system in which a band inversion occurs without spin-orbit interaction. For this purpose, f.c.c. SnSe is a good platform.

Although Ag$_{1-x}$Sn$_{1+x}$Se$_2$ is attracting attention as a superconductor\cite{9,10,11,11.5} and thermoelectric material\cite{11.5,12,13,14}, we focus on its topological properties. The results of an X-ray experiment\cite{9} show that the crystal structure of Ag$_{1-x}$Sn$_{1+x}$Se$_2$ is f.c.c. (space group $Fm\bar{3}m$, no. 225) for $0<x<0.24$. The relation between the lattice constant ($a_0$) and $x$ is given by $a_0=5.680+0.290x ~[\AA]$. For $x>0.24$, the f.c.c. structure is unstable. When $x=1$, i.e., in the case of SnSe, the crystal structure is known to be a TlI structure (space group $Cmnm$, no. 63), which is different from the f.c.c. structure. However, in order to focus on a Dirac nodal line, we assume the f.c.c. structure even for $0.24<x<1$.

In the following, we show the results for SnSe. We use a slightly smaller $a_0$ than the previous study ($a_0 = 5.99 \mathrm{\AA}$), but the result is consistent. The orbitals in each atom are determined by the pseudo potential.
We use the 4s, 4p, and 3d orbitals for Se, and the 5s, 5p, and 4d orbitals for Sn. Non-equivalent 6273 k-points in the first BZ are used in our band and density of states (DOS) calculations.
To confirm the accuracy of the calculations, we check the results with 897 k-points and find that the difference is negligible.


Figure 1(a) shows the total DOS without spin-orbit interaction.
Figures 1(b) and (c) show the partial DOS of the s, p, and d orbitals in Sn and Se, respectively. Figure 1(d) shows the band dispersion of f.c.c. SnSe, with the k-point path in the reciprocal lattice space shown in Fig. 2.
By comparing the partial DOS of Figs. 1(b) and (c) with the band dispersion, we find that three bands between -6.0 eV and 0 eV (Fermi energy) mainly originate from the Se 4p orbital, and three bands between 0 eV and 4.0 eV are mainly obtained from the Sn 5p orbital.

It is clear that there is a band crossing the W-L line.
As shown in Fig. 2, the W-L lines are located on the hexagonal parts of the BZ surface.
By considering the $O_h^5$ symmetry of the f.c.c. lattice, it is found that six band crossings exist on a hexagonal surface.

\begin{figure}
\begin{center}
\includegraphics[width=8.5cm]{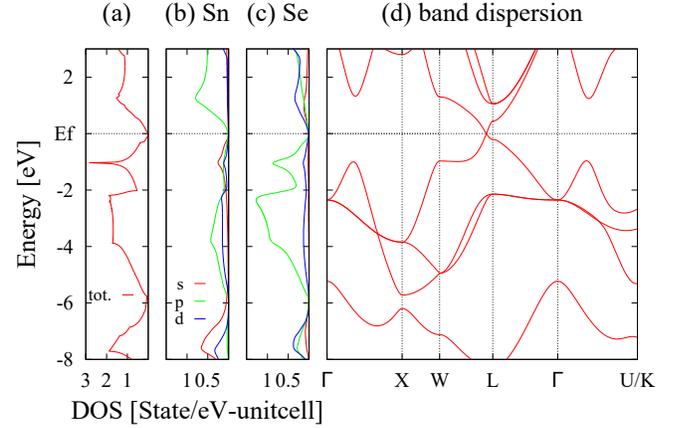}
\caption{(Color Online) 
(a) Total density of states (DOS) of SnSe. (b)(c) The partial DOS of the s (red line), p (green line), and d (blue line) orbitals in (b) Sn and (c) Se. (d) Band dispersion of f.c.c. SnSe. The k-point path in the reciprocal lattice space is shown in Fig. 2.}
\label{f1}
\end{center}
\end{figure}

\begin{figure}
\begin{center}
\includegraphics[width=5cm]{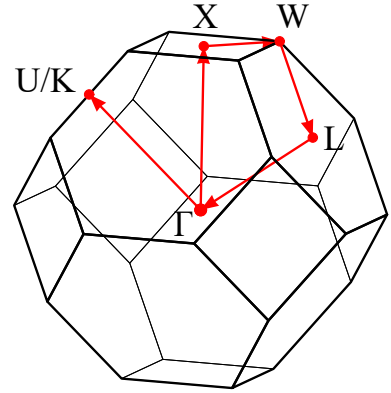}
\caption{(Color Online) Path of k-point in reciprocal lattice space.}
\label{f2}
\end{center}
\end{figure}

Next, we clarify the origin of the band crossing by studying the irreducible representation of the wave function for each band. 
Figure 3 shows the band dispersion around the L-point and the irreducible representation of each band.
The symbols $D_{2d}$, $C_2$, $D_{3d}$, and $C_{1h}$ on the upper side indicate the point groups of the corresponding k-points with Schoenflies notation.
The irreducible representations of the bands, $\mathrm{\Gamma}_1$, etc., are determined on the basis of these symmetries.

\begin{figure}
\begin{center}
\includegraphics[width=8.5cm]{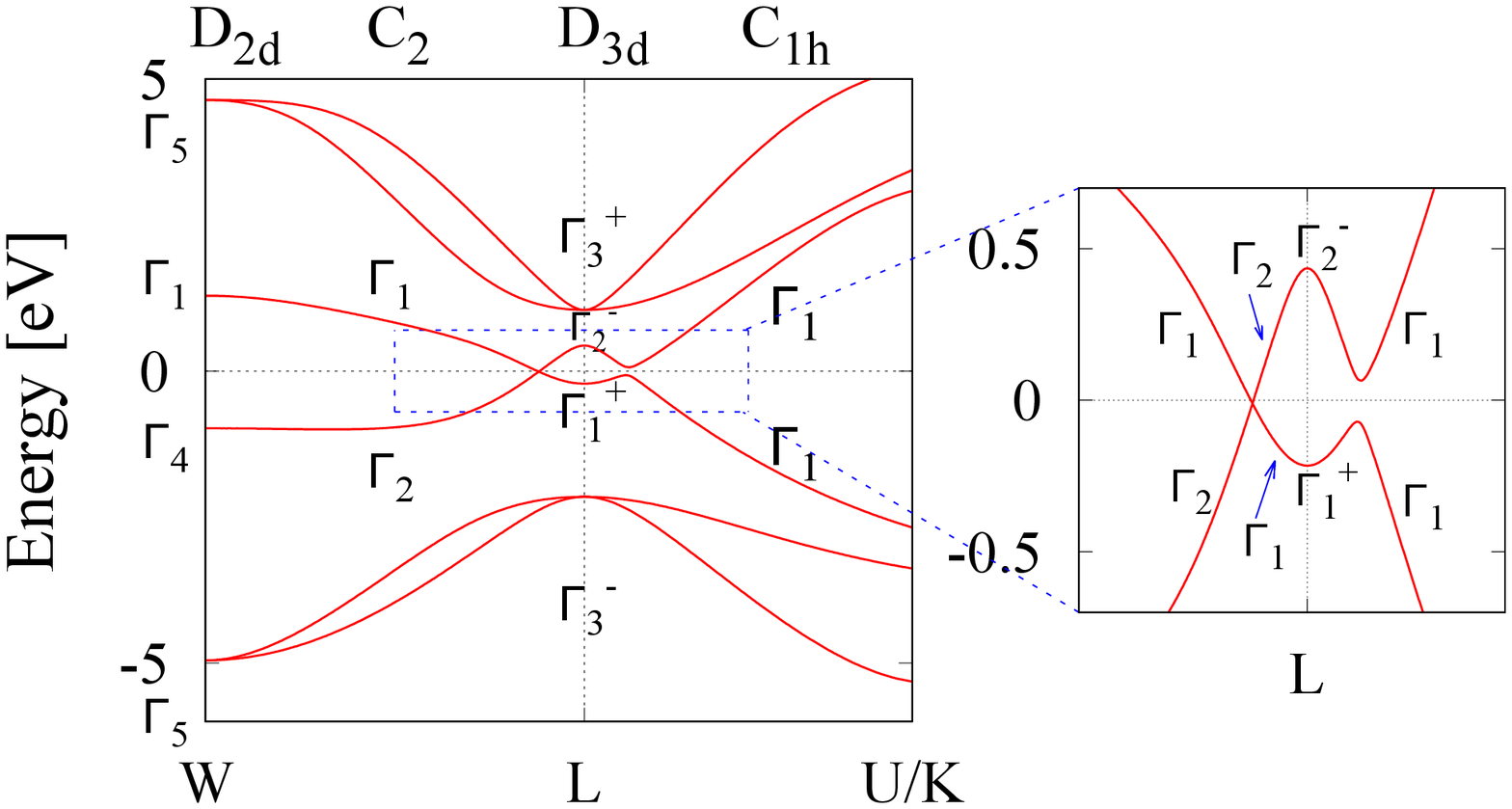}
\caption{(Color Online) Band dispersion around L-point without spin-orbit interaction. The irreducible representation of each band is also shown.
The characters ($D_{2d}$, $C_2$, $D_{3d}$, and $C_{1h}$) represent the point groups of the corresponding k-points. The right panel shows the expansion near the L-point.
}
\label{f3}
\end{center}
\end{figure}
In general, when two bands with different irreducible representations cross, there is no hybridization between them, and thus no band repulsion occurs. This happens on the L-W line, as shown in the right panel of Fig. 3. 
In contrast, on the L-U/K line, the irreducible representations of wave functions have the same $\Gamma_1$. Therefore, there is a small band repulsion (the right panel of Fig. 3). 

Let us discuss whether the band crossing on the L-W line is a point node or part of a line node.
As described in the introduction, in a system with the inversion and TR symmetries and without spin-orbit interaction, the point nodes cannot exist, and the band crossings must be a part of a nodal line~\cite{1,15}.
Because f.c.c. SnSe satisfies these conditions, the band crossing on the W-L line must be a part of a nodal line. Actually, we find that the Dirac nodal line exists at the Fermi level on a hexagonal surface of the BZ, as shown in Fig. 4. Because there are four non-equivalent hexagonal surfaces per BZ, there are four non-equivalent Dirac nodal lines per BZ, as shown in Fig. 4(a). Here, "non-equivalent" means that the two surfaces (Dirac nodal lines) are not connected by a reciprocal lattice vector.
As shown in Fig. 4(b), the nodal line penetrates the surface of the BZ right on the W-L line (shown by straight solid lines) and does not touch the L-U/K line (shown by straight dashed lines). Although the degenerate point on the W-L line is protected by $C_2$ symmetry, the Dirac nodal line itself is robust against $C_2$ symmetry breaking perturbation, as long as inversion symmetry exists. This is explained later in the discussion on the Berry phase.
\begin{figure}
\begin{center}
\includegraphics[width=8cm]{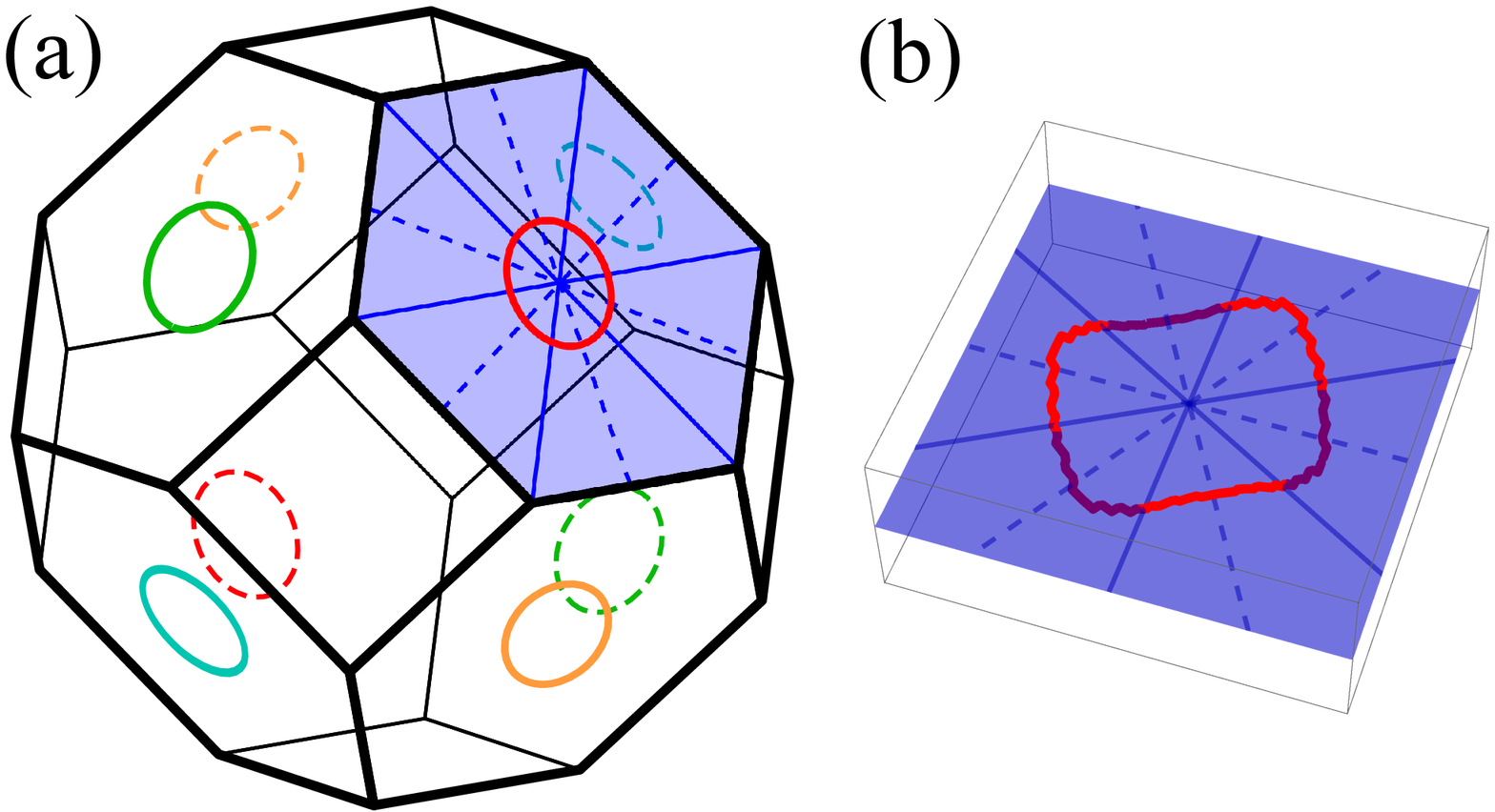}
\caption{(Color Online) All axis labels are wave numbers $k$: (a) BZ of f.c.c. and Dirac nodal lines (red, green, orange, and light blue circles). A dashed line is equivalent to a solid one with the same color. There are four non-equivalent Dirac nodal lines per BZ. (b)Details of the shape of the Dirac nodal line. The blue plane represents the hexagonal surface of BZ. The red line is the Dirac nodal line. The perpendicular to the surface has been magnified 10 fold. It penetrates the surface of the BZ right on the W-L line (solid line) and does not touch the L-U/K line (dashed line).}
\label{f5}
\end{center}
\end{figure}


We calculate the $\mathbb{Z}_2$ index for f.c.c. SnSe and find that the $\mathbb{Z}_2$ index is trivial, i.e., (0;000). 
Here, we explain why the Dirac nodal line exists even with a trivial $\mathbb{Z}_2$ index.
The eight TRIM in the BZ of the f.c.c. structure consist of one $\Gamma$-point, three X-points, and four L-points.
In SnSe, band inversion occurs around the L-points, and the nodal lines also appear around the L-points. Band inversion on an L-point changes the sign of the product of the inversion eigenvalues (see Table\ref{t1}).
Note that the $\mathbb{Z}_2$ indices for weak topological insulators $\nu_i ~(i=1,2,3)$ are defined as the products of four TRIM.
Those TRIM are composed of two X-points and two L-points, as listed in Table\ref{t2}. Because each index consists of an even number of L-points, i.e., 4 or 2, the sign of the inversion eigenvalue of the L-point has nothing to do with the $\mathbb{Z}_2$ index. In the case of $\mathbb{Z}_2$ non-trivial systems, Dirac nodal lines exists around band inverted points, and the band inversion changes the $\mathbb{Z}_2$ index. However, in f.c.c. SnSe, the band inversion occurs at even points, and thus the $\mathbb{Z}_2$ index is not changed.

\begin{figure}
\begin{center}
\includegraphics[width=6cm]{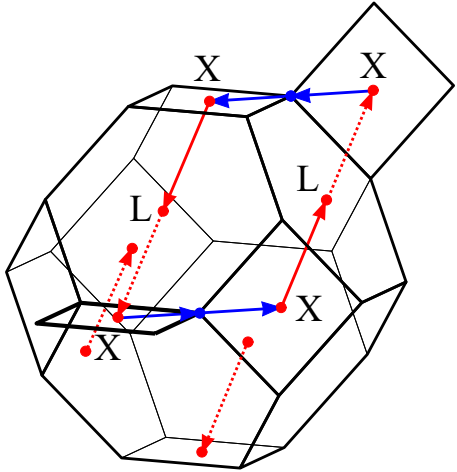}
\caption{(Color Online) Time-reversal invariant loop that corresponds to $\nu_1$. Parallelogram X, (L) X, X, (L) X is time-reversal invariant loop. The dashed arrows in the parallelogram exist outside the BZ, and an equivalent path in the BZ is also shown by isolated dashed arrows. The blue arrows are not time-reversal invariants. However, the two blue edges are equivalent, with opposite directions, so they cancel each other and do not contribute to the $\mathbb{Z}_2$ index.}
\label{f6}
\end{center}
\end{figure}

This trivial $\mathbb{Z}_2$ index can also be explained in terms of the time-reversal invariant loop.
An example of the TR invariant loop that corresponds to $\nu_i ~(i=1,2,3)$ is shown in Fig.\ref{f6}.
The TR loop goes through the two non-equivalent L-points. Here, "non-equivalent" means that the two L-points are not connected by a reciprocal lattice vector and each L-point is surrounded by a nodal line. Because the loop goes through the two non-equivalent L-points, we can understand that the TR loop is penetrated by two nodal lines. 
Thus, the Berry phase for this loop is $\pi+\pi = 2\pi$ or $\pi-\pi = 0$, and thus the $\mathbb{Z}_2$ index is trivial.

\begin{table}
\caption{Product of inversion eigenvalue of each TRIM without and with band inversion. "Without band inversion" means all of the irreducible representations of the valence band originate from the Se 4p orbital. "With band inversion" means the valence and conduction bands are inverted on the L-point, as can be seen in our results for SnSe.}
\begin{tabular}{|c|c|c|c|c|}
\cline{1-2} \cline{4-5}
\multicolumn{2}{|c|}{without band inversion} & & \multicolumn{2}{|c|}{with band inversion} \\
\cline{1-2} \cline{4-5}
point & \shortstack{inversion \\ eigenvalue} & & point & \shortstack{inversion \\ eigenvalue} \\
\cline{1-2} \cline{4-5}
$\Gamma$ & 1 & & $\Gamma$ & 1 \\
X & 1 & $\longrightarrow$ & X & 1 \\
L & -1 & & L & 1 \\
\cline{1-2} \cline{4-5}
\end{tabular}
\label{t1}
\end{table}

\begin{table}
\caption{Components of TRIM for each index. $\nu_0$ is written as a product of all eight TRIM. The other three $\nu_i$ are written as products of four TRIM: X, X, L, and L.}
\begin{tabular}{|c|c|}
\hline
$\mathbb{Z}_2$ index & calculated TRIM \\
\hline \hline
& $\Gamma$ \\
$\nu_0$ & X, X, X \\ 
& L, L, L, L \\
\hline
$\nu_1$ & \\
\cline{1-1}
$\nu_2$ & X, X, L, L \\
\cline{1-1}
$\nu_3$ & \\
\hline
\end{tabular}
\label{t2}
\end{table}

So far, we find that the TR loop is penetrated by two Dirac nodal lines, and thus they cannot be characterized by the $\mathbb{Z}_2$ index. 
Next, we check whether a single Dirac nodal line has topological properties.
Using an effective model, we will calculate the Berry phase for a loop that is penetrated by a single 
nodal line.
 First, we construct a $7 \times 7$ tight-binding model using the Slater--Koster's method\cite{16}, with the 5s and 5p orbitals for Sn and the 4p orbital for Se. Based on this tight-binding model, we derive a $2 \times 2$ effective model using the ${\bf k}\cdot{\bf p}$ perturbation.
As a result, we obtain an effective model for calculating the Berry phase given by
\begin{eqnarray}
H_{\text{eff}} &=& \left[ \epsilon + \lambda k_{\perp}^2 + \zeta (k_1^2 + k_2^2 ) \right] \sigma_0 \nonumber \\
& & + \left[ \epsilon' + \lambda' k_{\perp}^2 + \zeta' (k_1^2 + k_2^2 ) \right] \sigma_z \label{e1} \\
& & + \xi k_{\perp} \sigma_x, \nonumber
\end{eqnarray}
where $\sigma_0$ is an identical matrix, $k_{\perp}$ is the unit vector in the k-space with the [111] direction, and $k_1,k_2$ are the other two unit vectors perpendicular to $k_{\perp}$. The parameters are determined as follows: $\epsilon=-0.006, \lambda = 0.105, \zeta=-0.010, 
\epsilon'=0.214, \lambda' = -0.147, \zeta'=-0.126, and \xi = -0.634$.
For this $2 \times 2$ Hamiltonian written using the Pauli matrices, we can define d-vector as

\begin{eqnarray}
H_{\text{eff}} = d_0( {\bm k} ) \sigma_0 + \bm{d}({\bm k}) \cdot {\bm \sigma}.
\label{e3}
\end{eqnarray}
The eigenvalues of this Hamiltonian are given by $E(\bm{k})_{\pm}= d_0(\bm{k}) \pm | \bm{d}(\bm{k}) |$. The Dirac nodal line satisfies $E(\bm{k})_+ =E(\bm{k})_- \Leftrightarrow |\bm{d}(\bm{k})|=0$. By solving these equations, we can see that the Dirac nodal line exists on the line satisfying $k_{\perp}=0$ and $k_1^2 + k_2^2= |\epsilon'/\zeta'|$.

\begin{figure}
\begin{center}
\includegraphics[width=9cm]{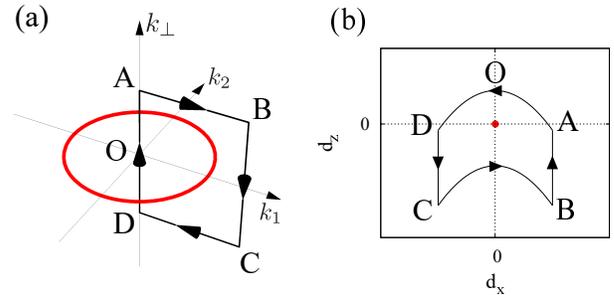}
\caption{(Color Online) (a)Nodal line (red) and path (black OABCD) for calculating Berry phase. (b)Path of d-vector in d-space, $d_x-d_z$ plane. Each point of OABCD corresponds to those of OABCD in (a). The origin (red point) corresponds to the nodal line. Path OABCD encircles the origin.}
\label{f7}
\end{center}
\end{figure}

Using this effective Hamiltonian, the Berry phase is easily calculated as a winding number of the d-vector\cite{17}.
We use a square loop, which is shown in Fig.\ref{f7}(a). The radius of the Dirac nodal line (red line) is $\sqrt{|\epsilon'/\zeta'|}$. We assume that $k_1$ is larger than $\sqrt{|\epsilon'/\zeta'|}$ on the BC edge in Fig.\ref{f7}(a). As a result, this loop is penetrated once by the Dirac nodal line.
From eq.\ (\ref{e1}), $\bm{d}(\bm{k})$ is calculated explicitly.
Because the coexistence of the TR and inversion symmetries requires $d_y(\bm{k})=0$, the d-vector is on the $d_x-d_z$ plane.
Because $\epsilon'>0$, $\lambda'<0$, and $\zeta'<0$ as a result of the {\bf k}$\cdot${\bf p} perturbation, the d-vector path for the loop OABCD is shown in Fig.\ref{f7}(b). Then, we can show that the parabolic curve BC in Fig.\ref{f7}(b) exists in the $d_z<0$ area.
Because the d-vector path surrounds the origin of the d-space, the winding number is one.
This calculation confirms that the single nodal line in our $\mathbb{Z}_2$ trivial system gives the $\pi$ Berry phase.
We also discuss the effect of the $C_2$ symmetry breaking perturbation. The $C_2$ symmetry breaking perturbation can change the d-vector. However, when inversion symmetry is kept, the d-vector is still confined on the $d_x-d_z$ plane, and the winding number is well-defined. Because a small perturbation cannot change the winding number, the Dirac nodal line and its $\pi$ Berry phase are robust against the $C_2$ symmetry breaking perturbation.

\begin{figure}
\begin{center}
\includegraphics[width=9cm]{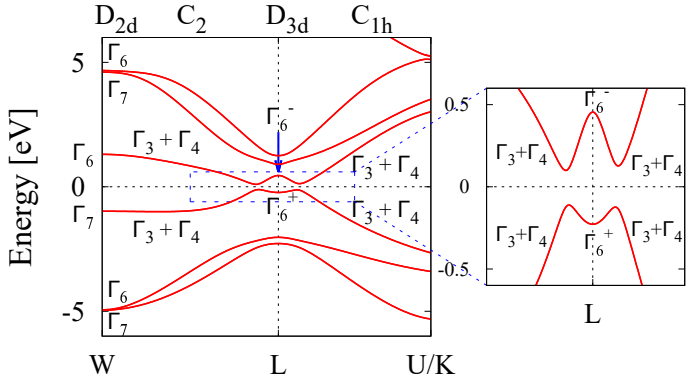}
\caption{(Color Online) Band dispersion around L-point with spin-orbit interaction and gap. The irreducible representation of each band is also shown. They are the same as those of the previous study\cite{8.5}.}
\label{f8}
\end{center}
\end{figure}

The qualitative effect of spin-orbit interaction is easily understood. The band dispersion with spin-orbit interaction is shown in Fig.\ref{f8}. We can see that the band dispersion is gapped, and its direct gap is approximately 0.2 eV. As described in the introduction, SnSe is known to be a topological crystalline insulator with a mirror Chern number of -2\cite{8.5}. Because of the irreducible representation and band gap, we can see that our result is consistent with that of the previous study.

\begin{figure}
\begin{center}
\includegraphics[width=7cm]{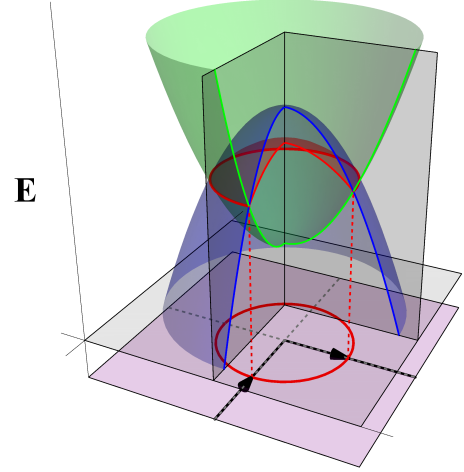}
\caption{(Color Online) Bulk states (green and blue) and surface state (red). Purple square shows reciprocal space on surface and red ring shows nodal line on it. In fact, there are many bulk states, and two band directly making nodal lines are selected. Surface state (red) exists inside nodal line.}
\label{f9}
\end{center}
\end{figure}

A surface state can be derived using the effective model of eq.\ (\ref{e1}). First, we neglect the $\sigma_0$ term because it only shifts the energy eigenvalues.
To discuss the surface state, we focus on the (111) surface.
Then, we can replace $k_{\perp}$ with $-i \partial_{\perp}$, and we assume that $r_\perp = +\infty$ is outside of the crystal. The $k_\perp^2$ term requires a more complicated treatment. We replace it with a smooth function $\Delta(r_\perp)$, which satisfies the following conditions. Comparing the sign of the $\sigma_z$ term from the bulk Hamiltonian (\ref{e1}), $\Delta (r_\perp)$ must satisfy $\Delta (r_\perp)=0$ for $r_\perp = - \infty$ and $\Delta (r_\perp)<-\epsilon'$ for $r_\perp = + \infty$.
As a result, an effective Hamiltonian is written as follows:
\begin{eqnarray}
H_{\text{sur}} = -i \xi \partial_{\perp} \sigma_x + \left[ \Delta(r_\perp) + \epsilon' + \zeta' (k_1^2+k_2^2) \right] \sigma_z.
\label{e4}
\end{eqnarray}
The Hamiltonian of eq. (\ref{e4}) is well known as the Jackiw--Rebbi problem\cite{18}. It is known that an eigenstate with $E=0$ exists, which is localized around the point satisfying the condition, $\Delta(r_\perp)+\epsilon'+\zeta' (k_1^2+k_2^2)=0$.
By definition, this point exists around the surface, and we can see that this localized state is the surface state.
Because this surface state requires the existence of a point that satisfies the condition of $\Delta(r_\perp)+\epsilon'+\zeta' (k_1^2+k_2^2)=0$, this surface state emerges inside the nodal line (Fig.\ref{f9}).
This kind of surface state is sometimes called a "drumhead"-like surface state and is a typical surface state of a system with Dirac nodal lines.

In conclusion, we found by a first principles calculation that a Dirac nodal line emerges in f.c.c. SnSe, which has a trivial $\mathbb{Z}_2$ index in the absence of spin-orbit interaction.
The trivial $\mathbb{Z}_2$ index comes from the fact that even L-points are contained in the BZ. We also showed that the single nodal line gives the $\pi$ Berry phase and a "drumhead"-like surface state exists, as generally expected for a Dirac nodal line. The situation is similar to the emergence of a topological crystalline insulator.

However, the relation between the mirror Chern number and nodal line in f.c.c. SnSe was not revealed in this letter. The same situation, i.e., an even number of high symmetry points with the same symmetry in the BZ, can give a Dirac nodal line in the $\mathbb{Z}_2$ trivial system. For example, a body centered tetragonal system such as space group no. 139 is one of the candidates. Finding easy methods to find the Dirac nodal lines in $\mathbb{Z}_2$ trivial systems and classify them remains as future work.



\begin{acknowledgment}
We acknowledge the many fruitful discussions with Masao Ogata, Toshikaze Kariyado, and Kaya Kobayashi. This work was supported by the JSPS Core-to-Core Program, A. Advanced Research Networks. We were also supported by Grants-in-Aid for Scientific Research from the Japan Society for the Promotion of Science (Nos. 15K17694, 25220803, 15H02108, 17H02912, 17H02923, 18K03482, and 18H01162 ). I.T. was supported by the Japan Society for the Promotion of Science through the Program for Leading Graduate Schools (MERIT).

\end{acknowledgment}


\begin{thebibliography}{10}
\bibitem{1} Y. Kim, B. J. Wieder, C. L. Kane, and A. M. Rappe, Phys. Rev. Lett. $\bm{115}$, 036806 (2015).
\bibitem{2} Y. Du, F. Tang, D. Wang, L. Sheng, E. Kan, C.-G. Duan, S. Y. Savrasov, and X. Wan, Njp Quantum. Matter. $\bm{2}$, 3 (2017).
\bibitem{3} L. S. Xie, L. M. Schoop, E. M. Seibel, Q. D. Gibson, W. Xie, and R. J. Cava, APL Mater. $\bm{3}$, 083602 (2015).
\bibitem{3.05} M. Hirayama, R. Okugawa, T. Miyake, and S. Murakami, Nat. Commun. $\bm{8}$, 14022 (2017).
\bibitem{3.1} M. Neupane, I. Belopolski, M. M. Hosen, D. S. Sanchez, R. Sankar, M. Szlawska, S.-Y. Xu, K. Dimitri, N. Dhakal, P. Maldonado, P. M. Oppeneer, D. Kaczorowski, F. Chou, M. Z. Hasan, and T. Durakiewicz, Phys. Rev. B $\bm{93}$, 201104(R) (2016).
\bibitem{3.2} B. Feng, B. Fu, S. Kasamatsu, S. Ito, P. Cheng, C.-C. Liu, Y. Feng, S. Wu, S. K. Mahatha, P. Sheverdyaeva, P. Moras, M. Arita, O. Sugino, T.-C. Chiang, K. Shimada, K. Miyamoto, T. Okuda, K. Wu, L. Chen, Y. Yao, and I. Matsuda,. Nat. Commun. $\bm{8}$, 1007 (2017).
\bibitem{4} S.-Y. Yang, H. Yang, E. Derunova, S. S. P. Parkin, B. Yan, and M. N. Ali, Adv. Phys. X $\bm{3}$,1414631 (2018).
\bibitem{5} L. Fu, C. L. Kane, and E. J. Mele, Phys. Rev. Lett. $\bm{98}$, 106803 (2007).
\bibitem{15} H. Huang, J. Liu, D. Vanderbilt, and W. Duan, Physical Review B, $\bm{93}$ 201114(R), (2016).
\bibitem{6} L. Fu, Phys. Rev. Lett. $\bm{106}$, 106802 (2011).
\bibitem{7} T. H. Hsieh, H. Lin, J. Liu, W. Duan, A. Bansil, and L. Fu, Nat. Commun. $\bm{3}$, 982 (2012).
\bibitem{8} Y. Tanaka, Z. Ren, T. Sato, K. Nakayama, S. Souma, T. Takahashi, K. Segawa, and Y. Ando, Nat. Phys. $\bm{8}$, 800 (2012).
\bibitem{8.5} Y. Sun, Z. Zhong, T. Shirakawa, C. Franchini, D. Li, Y. Li, S. Yunoki, and X.-Q. Chen, Phys. Rev. B $\bm{88}$, 235122 (2013).
\bibitem{qe} P. Giannozzi, S. Baroni, N. Bonini, M. Calandra, R. Car, C. Cavazzoni, D. Ceresoli, G. L. Chiarotti, M. Cococcioni, I. Dabo, A. Dal Corso, S. Fabris, G. Fratesi, S. de Gironcoli, R. Gebauer, U. Gerstmann, C. Gougoussis, A. Kokalj, M. Lazzeri, L. Martin-Samos, N. Marzari, F. Mauri, R. Mazzarello, S. Paolini, A. Pasquarello, L. Paulatto, C. Sbraccia, S. Scandolo, G. Sclauzero, A. P. Seitsonen, A. Smogunov, P. Umari, and R. M. Wentzcovitch, J. Phys. Condens. Matter $\bm{21}$, 395502 (2009).
\bibitem{52} J. P. Perdew, K. Burke, and M. Ernzerhof, Phys. Rev. Lett. $\bm{80}$, 891 (1998).
\bibitem{8.1} Y. W. Tung and M. L. Cohen, Phys. Rev. $\bm{180}$, 823 (1969).
\bibitem{9} D. C. Johnston and H. Adrian, J. Phys. Chem. Solids $\bm{38}$, 355 (1977).
\bibitem{10} Z. Ren, M. Kriener, A. A. Taskin, S. Sasaki, K. Segawa, and Y. Ando, Phys. Rev. B $\bm{87}$, 064512 (2013).
\bibitem{11} Y. Hijikata, A. Nishida, K. Nagasaka, O. Miura, A. Miura, C. Moriyoshi, Y. Kuroiwa, and Y. Mizuguchi, J. Phys. Soc. Jpn. $\bm{86}$, 054711 (2017).
\bibitem{11.5} T. Wakita, E. Paris, K. Kobayashi, K. Terashima, M. Y. Hacisaliho\u{g}lu, T. Ueno, F. Bondino, E. Magnano, I. P\'i\u{s}, L. Olivi, J. Akimitsu, Y. Muraoka, T. Yokoya, and N. L. Saini., Phys. Chem. Chem. Phys. $\bm{19}$, 26672 (2017).
\bibitem{12} S. Sassi, C. Candol, J.-B. Vaney, V. Ohorodniichuk, P. Masschelein, A. Dauscher, and B. Lenoir, Appl. Phys. Lett. $\bm{104}$, 212105 (2014).
\bibitem{13} C.-L. Chen, H. Wang, Y.-Y. Chen, T. Daya, and G. Jeffrey Snyder, J. Mater. Chem. A, 11171 (2014).
\bibitem{14} G. Tan, L.-D. Zhao, F. Shi, J. W. Doak, S.-H. Lo, H. Sun, C. Wolverton, V. P. Dravid, C. Uher, and M. G. Kanatzidis, Am. Chem. Soc. $\bm{136}$, 7006 (2014).
\bibitem{16} P. C. Slater and G. F. Koster, Phys. Rev. $\bm{94}$, 1498 (1954).
\bibitem{17} For example, J.K. Asb\'oth, L. Oroszl\'any, and A. P\'alyi, A Short Course on Topological Insulators (Springer, 2016).
\bibitem{18} R. Jackiw and C. Rebbi, Phys. Rev. D $\bm{13}$, 3398 (1976).
\end{thebibliography}
\end{document}